# Space-Time Phononic Crystals with Anomalous Topological Edge States

Mourad Oudich[1,2,*], Yuanchen Deng[1], Molei Tao[3], and Yun Jing[1,*]

[1]*Department of Mechanical and Aerospace Engineering, North Carolina State University, Raleigh, North Carolina 27695, USA*

[2]*Université de Lorraine, CNRS, Institut Jean Lamour, F-54000 Nancy, France*

[3]*School of Mathematics, Georgia Institute of Technology, Atlanta, Georgia 30332, USA*

\* moudich@ncsu.edu, yjing2@ncsu.edu

**Abstract**

It is well known that an interface created by two topologically distinct structures could host nontrivial edge states that are immune to defects. In this letter, we introduce a one-dimensional space-time phononic crystal and study the associated anomalous topological edge states. A space-decoupled time modulation is assumed. While preserving the key topological feature of the system, such a modulation also duplicates the edge state mode across the spectrum, both inside and outside the band gap. It is shown that, in contrast to conventional topological edge states which are excited by frequencies in the Bragg regime, the time-modulation-induced frequency conversion can be leveraged to access topological edge states at a deep subwavelength scale where the entire phononic crystal size is merely 1/5.1 of the wavelength. This remarkable feature could open a new route for designing miniature devices that are based on topological physics.

There has been a growing interest in topological phases since the discovery of topological insulators in quantum physics. One of the most intriguing phenomena associated with topological phases is the topological edge state, which is symmetry-protected and is immune to local material imperfections. Topological phases have also recently been identified in classical systems such as in photonics [1-11], acoustics [12-28], and mechanics [29-33]. These topological phases harbor nontrivial wave modes that are stable against local perturbations, offering an ingenious solution for realizing robust, backscattering-immune wave propagation [12-23] as well as energy localization [24-28]. While most recent studies pertaining to topological edge states have focused on two-dimensional (2D) [12-20,27-33] or even three-dimensional (3D) systems [21-23], one-dimensional (1D) systems [24-26] could still serve as a simple yet versatile platform for advancing our fundamental knowledge on topological physics. For the 1D case, the topological phase (or geometrical phase) of the bulk band is specifically known as the Zak phase [34,35]. Considering a design of periodic structure with a band gap (BG), it is discovered that the continuous deformation of the structure with constant symmetry can produce closing and opening of the BG supported by a topological phase transition. This phase transition can be leveraged to design topologically protected edge states that reside between the initial structure and a deformed one. Xiao *et. al* [24] demonstrated this mechanism in a 1D acoustic system where a topological edge state was created between two phononic crystals (PC) having topologically distinct BGs. Meanwhile, topological edge states are dictated by the geometrical symmetry of the system and occur at the topological BG where the wavelength is comparable with the structure periodicity [17]. This limits their functionalities to the Bragg scattering regime. One critical consequence, for instance, is that the resulting structure can be very bulky at low frequencies. This letter introduces a new strategy to break down this barrier by imposing both space and time modulation (STM) on the medium properties. The STM could enable the engineering of topological phases in the time domain which provides an extra degree of freedom and therefore richer physics. There are two major findings in this study: (1) the creation of anomalous topological edge states in the bulk band; (2) the possibility to excite the edge state in the Bragg regime using ultralow frequencies whose wavelengths could be orders of magnitude greater than the periodicity of the PC. These features could bring about new designs of miniature devices based on topological phases for unconventional wave manipulation, excitation, and detection. Although the current study focuses on 1D acoustic systems, the theory developed here can readily be extended to higher dimensions and other wave-based systems.

STM of the medium properties has been proposed as a robust way to break reciprocity which is an intrinsic property of wave dynamics in conventional systems. Consequently, non-identical wave transmission in opposite directions has been achieved. Space-time (ST) photonic crystals were in fact introduced decades ago [36-38] and the subject on non-reciprocity has recently been reinvigorated in the field of photonics for the control of electromagnetic waves using time-variation of permittivity and/or permeability [39-43]. There has also been numerous works regarding STM in acoustics [44,46] and elastic waves where the mechanical properties of the medium are ST modulated to achieve nonreciprocal wave propagation [47-50]. Meanwhile, it has been shown that external time-dependent perturbations can be used to achieve topological spectra in initially trivial systems at the equilibrium [6,16,51]. This class of systems is known to be the basis of Floquet topological insulators where time-periodic potential is introduced to a Hermitian system which leads to nontrivial duplicated edge states in the band structure [52-54].

In this work, in contrast to the approach for building Floquet topological insulators, we start from a stationary 1D system possessing nontrivial topological phases as well as topologically protected edge states inside a BG. This initial system is constructed from two space-modulated PCs exhibiting distinct Zak phases, i.e., 0 and π. We then introduce the time modulation into both PCs in a way that the Zak phase difference between the two PCs remains unchanged throughout the entire duration of time modulation [53]. The resulting PC is referred to as the space-time PC: a "living" PC that evolves in time. As will be shown below, frequency conversion occurs in the space-time PC, which gives rise to the duplication of edge states across the entire temporal spectrum. This, as shown later, is the underlying physics for the anomalous edge state discovered in this study.

We first consider two PCs having a sinusoidal space modulation of the medium density $\rho(x)$. The governing equations in acoustics in 1D read [55]:

$$\begin{aligned}\frac{\partial p}{\partial x} &= -\rho \frac{\partial v}{\partial t} \\ \frac{\partial v}{\partial x} &= -K_e \frac{\partial p}{\partial t}\end{aligned} \quad (1)$$

where $p$, $v$, $\rho$ and $K_e$ are the pressure, the particle velocity, the density, and the compressibility of the medium. This study assumes $K_e$ to be a constant, i.e., space-and-time independent, though we can conduct a similar study where the density is constant and $K_e$ is space-and-time dependent.

The top left panels of Figs.1(a) and (b) illustrate the unit cell for each PC: one has a modulation in the form of $\rho = \rho_0(1 - \alpha.\cos(k_m x))$ and the other is $\rho = \rho_0(1 + \alpha.\cos(k_m x))$ in the spatial domain $[0, \lambda_m]$, where $\lambda_m = 2\pi/k_m$ is the space period, $\rho_0$ is the air density (1.2 kg/m³), and $\alpha$ is the modulation amplitude (chosen as $\alpha = 0.5$ throughout this study). The two PCs are exactly 180° out of phase in space. Both PCs have identical band structures (BSs) with a BG around the normalized frequency Ω=0.5 between the first and second bands (bottom left panels of Figs.1 (a) and (b)). For each PC, we plot in the right panels of Figs.1(a) and (b) the acoustic pressure distributions along the unit cell at the edge of the Brillouin zone (BZ) in the first band (point 1), and at both the edge and center of the BZ in the second band (points 2 and 2'). The symmetry of the eigenmode dictates the Zak phase of each band. For instance, for the first PC in Fig.1 (a), points 2 and 2' in the second band have the same symmetry, indicating a zero Zak phase; for the second PC in Fig. 2 (b), however, different symmetries (anti-symmetrical vs. symmetrical) are present, implying that the Zak phase is $\pi$. The Zak phase can be also confirmed by using the following formula [24]:

$$\theta_Z = \int_{-\pi/\lambda_m}^{\pi/\lambda_m} \langle u_k | \partial_k | u_k \rangle dk = \frac{1}{2\rho_0 c_0^2} \int_{-\pi/\lambda_m}^{\pi/\lambda_m} \left( i \int_0^{\lambda_m} u_k^*(x) \partial_k u_k(x) dx \right) dk, \quad (2)$$

where $k$ is the wave number and $u_k = e^{-ikx} p$.

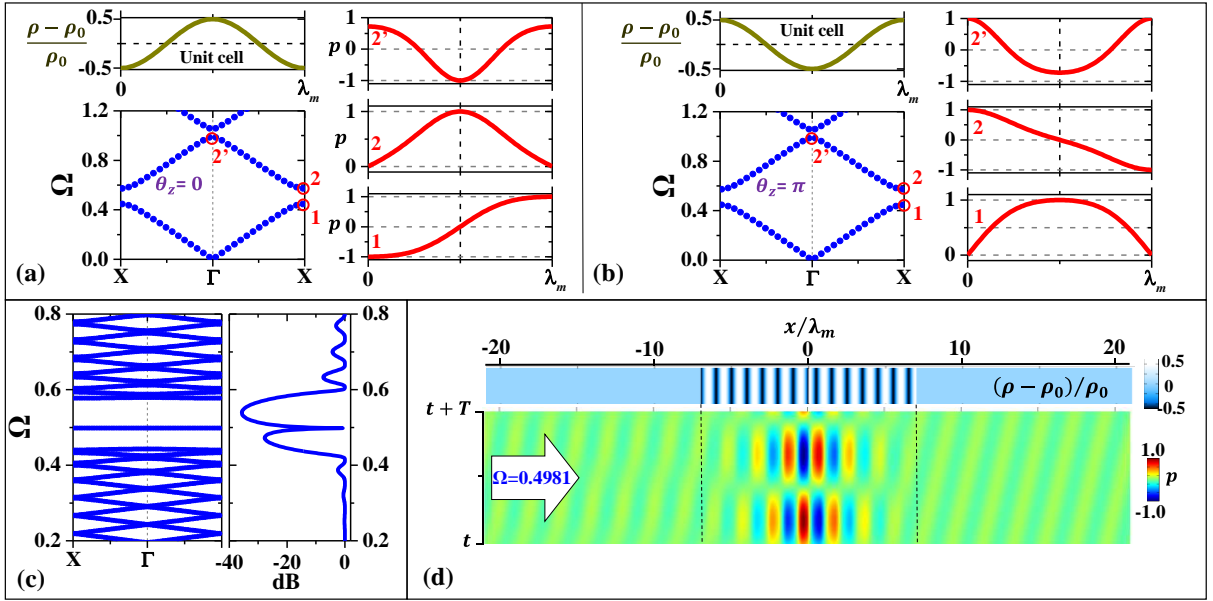

Fig1. (a) A stationary PC with density modulation of $\rho = \rho_0(1 - \alpha.cos(k_m x))$ along the unit cell $[0, \lambda_m]$ and its band structure (bottom left panel). Pressure fields are also plotted for the eigenmodes indicated by 1, 2 and 2' in the band structure (right panels). (b) Same as (a) but for the second stationary PC with density modulation of $\rho = \rho_0(1 + \alpha.cos(k_m x))$. (c) Band structure for the system constructed by the two PCs where each PC consists of 7 unit cells: $\rho = \rho_0(1 - \alpha.cos(k_m x))$ for $x < 0$ and $\rho = \rho_0(1 + \alpha.cos(k_m x))$ for $x > 0$. The corresponding transmission spectrum is shown on the right, where an edge state with a high transmission can be observed at $\Omega=0.4981$. (d) (Top panel) Transmission configuration along the system and (bottom panel) the edge state pressure field for a single frequency excitation plotted against the time. In both panels, the vertical axis is the time and the horizontal axis is the space.

Taking advantage of the Zak phase change in the second band, a topological edge state located at $x = 0$ using both PCs is constructed. The resulting system has a space modulation of $\rho = \rho_0(1 - \alpha.cos(k_m x))$ for $x < 0$ and $\rho = \rho_0(1 + \alpha.cos(k_m x))$ for $x > 0$. The BS and the transmission spectrum of this system are plotted in Fig.1(c) which both show the existence of an edge state. The calculations are performed in the frequency domain using the commercial software Comsol Multiphysics® v5.3a. This edge state is located at the middle of the Bragg BG, i.e., $\Omega = \omega/k_m c_0 = 0.4981$, at which the structure periodicity corresponds to half of the wavelength. We plot in the top panel of Fig.1 (d) the density distribution of the system in a transmission configuration where each PC consists of 7 unit cells (14 in total). For a monochromatic wave excitation at the edge state frequency $\Omega = 0.4981$, the pressure field distribution (real part of the complex pressure) is plotted in Fig.1 (d) during one time period $T = 2\pi/\omega$. The acoustic energy is clearly confined in the vicinity of the interface between the two PCs (*x*=0). It is noted that although sinusoidal space modulation is chosen in this study, the theory developed here can be generalized to other types of modulation (e.g., square modulation) and the main physics is expected to be the same.

The time modulation is subsequently introduced to both PCs in the following manner:

$$\rho = \rho_0\big(1 - (\alpha + \beta \sin(\omega_m t)).\cos(k_m x)\big) \qquad \text{for} < 0,$$
$$\text{and} \quad \rho = \rho_0\big(1 + (\alpha + \beta \sin(\omega_m t)).\cos(k_m x)\big) \qquad \text{for} > 0. \qquad (3)$$

where $\alpha = 0.5$ and $\beta = 0.2$ are assumed. A modulation frequency of $\Omega_m = \omega_m/k_m c_0 = 0.484$ is chosen which is close to the edge state's frequency of the stationary system (corresponding to $\beta = 0$). This allows us to excite an ultralow frequency edge state mode as will be shown later. Please note that the density here is not assumed to be weakly modulated as was done in [45]. This time modulation is space-decoupled in the sense that the functions (sine and cosine) governing the space and time modulations are independent from each other, which is largely unexplored in the past. One can visualize this as the density of the system only having its amplitude varying with time. Fig. 2 (a) shows the time variation of the reduced density $(\rho - \rho_0)/\rho_0$ of the super-cell having 14 unit cells. This constitutes the domain in which Eq. 1 is to be solved. Another type of ST modulation where the density has a wave-like variation (so-called space-coupled time modulation), which has been explored for non-reciprocal propagation, is discussed in the supplementary material where the fundamental differences between the two modulation are delineated [56]. With the space-decoupled time modulation of the super cell described by Eq. (3), each PC has its Zak phase fixed with time. This is because only the amplitude of the density is changing with time, so that the symmetry of the second band is conserved at each instant. Consequently, the Zak phase difference between the two PCs remains independent of time.

Knowing that the density $\rho$ is periodic in both space and time, one can apply the Floquet theorem:

$$p(x,t) = \tilde{p}(x,t)e^{i(\omega t - kx)}$$
$$v(x,t) = \tilde{v}(x,t)e^{i(\omega t - kx)} \qquad (4)$$

where $\tilde{p}$ and $\tilde{v}$ are periodic both in time and space. Inserting these expressions into Eq. (1), the following eigenvalue problem can be constructed where $\omega$ is the eigenvalue to be solved for each wavenumber $k$:

$$\begin{pmatrix} \frac{\partial}{\partial x} & \rho(x,t)\frac{\partial}{\partial t} \\ K_e \frac{\partial}{\partial t} & \frac{\partial}{\partial x} \end{pmatrix}\begin{pmatrix}\tilde{p}\\\tilde{v}\end{pmatrix} + \begin{pmatrix} -ik & 0 \\ 0 & -ik \end{pmatrix}\begin{pmatrix}\tilde{p}\\\tilde{v}\end{pmatrix} = \omega \begin{pmatrix} 0 & -i\rho(x,t) \\ -iK_e & 0 \end{pmatrix}\begin{pmatrix}\tilde{p}\\\tilde{v}\end{pmatrix} \qquad (5)$$

The time variable *t* can be considered as a synthetic dimension in addition to the real physical dimension *x*. Subsequently, Eq. (5) can be solved in the space-time domain displayed in Fig. 2(a) along with the Floquet boundary conditions. Such a method can be generalized to solving wave equation problems involving media with arbitrary space-time periodic properties.

In this study, Eq. (5) is solved again by Comsol Multiphysics®. Among all possible solutions, we have identified three eigen-modes displayed in Figs 2.(b)-(d) where the acoustic energy is confined within the vicinity of *x*=0: the first one is located at frequency $\Omega_0 = 0.4981$ (Fig 2.(c)) and it corresponds to the edge state inside the BG in the stationary case; the second and third modes are located exactly at $\Omega_0 + \Omega_m = 0.9821$ (Fig 2.(b)) and $\Omega_0 - \Omega_m = 0.014$ (Fig 2. (d)). These modes are anomalous edge state, and their emergence in the bulk band can be attributed to the time-modulation-induced frequency conversion, which changes the topological feature of the structure leading to duplicated edge states at frequencies $\Omega_0 \pm n\Omega_m$ where $n$ is an integer [52-54]. The positions of these modes in the band can be tuned by changing the modulation frequency $\Omega_m$ as shown in Fig 2. (e), where the theory (lines) agrees very well with the simulation (circles).

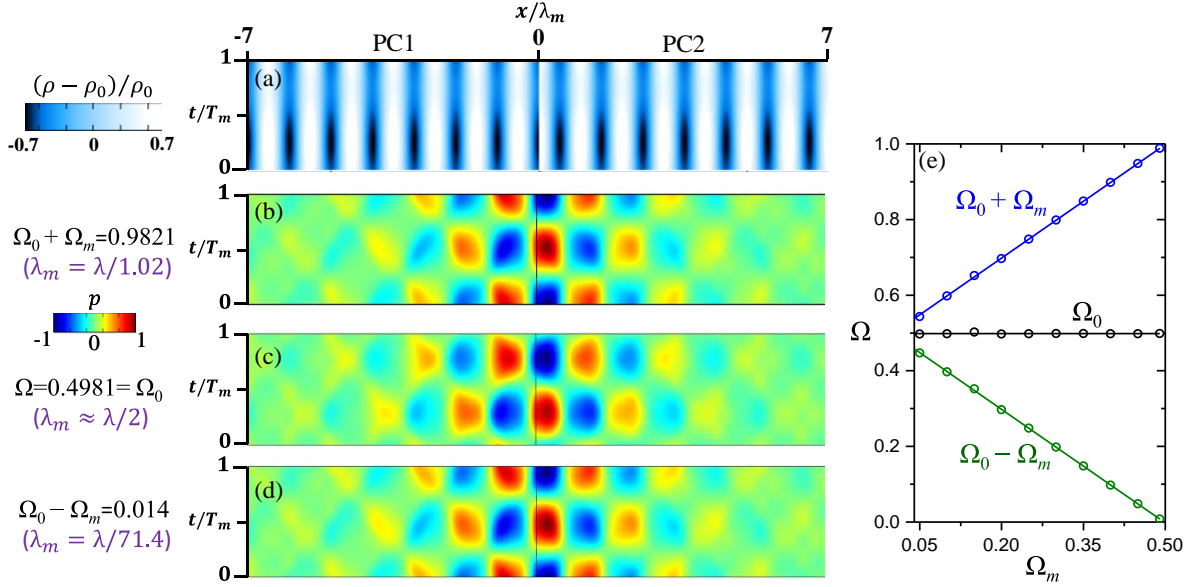

Fig.2. (a) Density modulation of $\rho = \rho_0\big(1 - (\alpha + \beta sin(\omega_m t)).cos(k_m x)\big)$ for $x < 0$ and $\rho = \rho_0\big(1 + (\alpha + \beta sin(\omega_m t)).cos(k_m x)\big)$ for $x > 0$ presented within one modulation period $T_m$. The wave fields of the edge state (c) inside the BG and duplicated edge states located at $\Omega_0 + \Omega_m$ (b) and $\Omega_0 - \Omega_m$ (d). (e) Frequency of the edge states as function of the modulation frequency $\Omega_m$. Lines are theoretical prediction and circles are simulation results.

To gain a better insight on these modes, Fig. 3(a) illustrates the calculated transmission spectrum through the STM system. The calculation is performed in the time domain where we analyze the Fourier spectrum of the transmitted wave through the structure using a wideband incident pulse (a Gaussian-modulated cosine function). A transmission peak corresponding to the edge state is observed inside this BG at the normalized frequency $\Omega_0 = 0.4981$ which corresponds exactly to the stationary case. Interestingly, the system remains reciprocal and the BG does not experience frequency shift from the stationary case, which is a major departure from the commonly studied space-coupled time modulation where reciprocity is broken [45-47]. In these systems, the space-time perturbation or modulation of the intrinsic property of the medium (density, compressibility or stiffness) has a driven or wavelike directional motion which breaks the reciprocity. This is not the case for our space-decoupled time modulation which preserves reciprocity. We also observe two less visible transmission peaks at 0.014 and 0.9821 (insets of Fig. 3(a)) corresponding to the excitation of the first-order edge states outside the BG at $\Omega_0 \pm \Omega_m$, which confirms the prediction of the earlier eigenvalue analysis. These peaks appear less pronounced in the spectrum compared to the one at $\Omega_0$ because they are located outside the BG. The bottom panel of Fig. 3(b) shows the wave pressure field for a monochromatic excitation at the frequency $\Omega_0 = 0.4981$ where one can observe the localized edge state mode at the interface $x$=0. This mode, however, is not pure in its Fourier components, although the excitation is strictly monochromatic. To this end, we apply temporal Fourier transform to the wave field shown in Fig. 3(b) and the results (referred to as the spatial-frequency spectrum of the wave field) are presented in Fig. 3(c) for the case of monochromatic excitations at $\Omega_0 = 0.4981$ and $\Omega_0 - \Omega_m = 0.014$, respectively plotted in the left and right panels. In other words, these are the pressure fields along the PC across the frequency spectrum. The spatial domain has a range of $-20 < x/\lambda_m < 20$. In the case of an incident wave at $\Omega_0 = 0.4981$ (left panel of Fig. 3(c)), edge state modes confined at the center of the system (in the vicinity of $x$=0) are excited both at this same frequency and at $\Omega_0 - \Omega_m = 0.014$, albeit with the latter having a much smaller amplitude. Meanwhile, if we choose a monochromatic excitation exactly at $\Omega_0 - \Omega_m = 0.014$, the wave field plotted in the right panel of Fig. 3(c) suggests that it is possible to excite the edge state at $\Omega_0 = 0.4981$ inside the BG. We also would like to point out that it is likely that the edge state at $\Omega_0 - \Omega_m$ is also excited but its characteristics (e.g., energy confinement) is not clearly visible as it is "drown" by the wave field excitation at this frequency. This has happened partially due to the fact that this frequency, again, is located outside the BG.

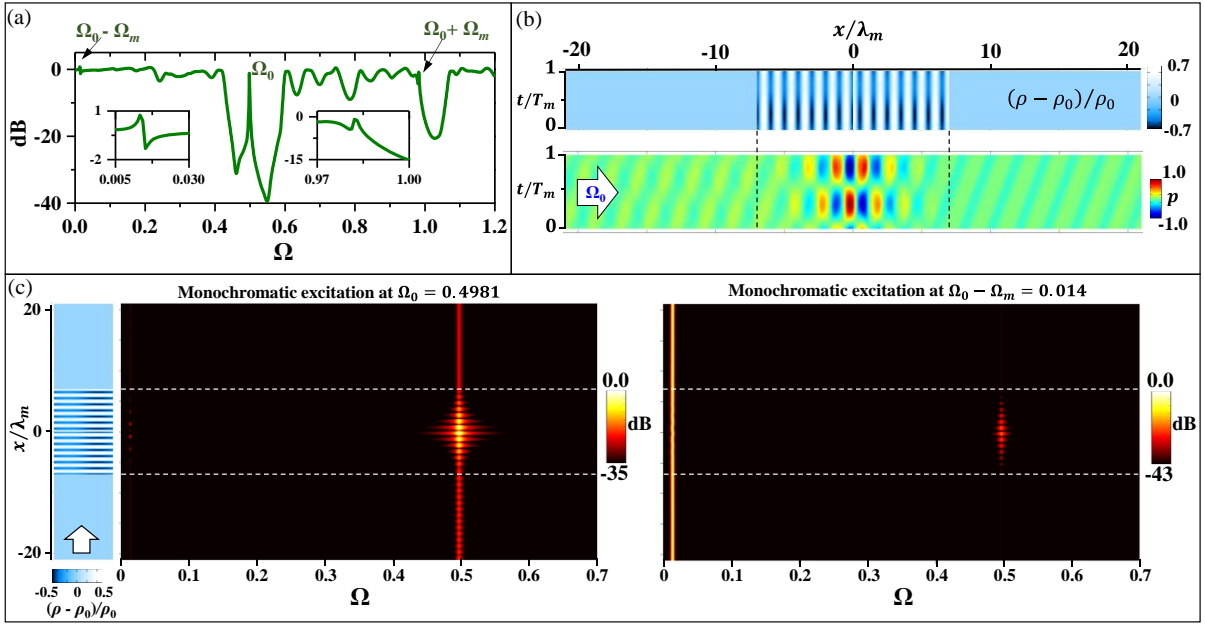

Fig.3. (a) Transmission spectrum through the space-decoupled time modulated system composed of two PCs with a constant Zak phase shift, and a close-up view near $\Omega_0 \pm \Omega_m$ (inset figures). (b) (Top panel) Transmission configuration of the structure with density modulation of $\rho = \rho_0(1 - (\alpha + \beta sin(\omega_m t)).cos(k_m x))$ for $x < 0$ and $\rho = \rho_0(1 + (\alpha + \beta sin(\omega_m t)).cos(k_m x))$ for $x > 0$ presented within one modulation period $T_m$. (Bottom panel) Pressure field for incident wave at $\Omega_0 = 0.4981$. (c) Spatial-frequency spectra of the wave field for incident wave excitations at $\Omega_0 = 0.4981$ (left panel) and $\Omega_0 - \Omega_m = 0.014$ (right panel).

Remarkably, for the case of monochromatic excitation at $\Omega_0 - \Omega_m = 0.014$, the incident wavelength is at $\lambda \approx 71.4\, \lambda_m$ and the frequency conversion allows the excitation of the edge state inside the BG at $\Omega_0 = 0.4981$ in the Bragg regime. Giving that the entire phononic crystal has a length of $L_s = 14\lambda_m$, the functionality of the structure is effective though its size is only 1/5.1 of the incident wave wavelength (vs. 7×wavelength in the stationary case). Though the peak pressure of the first-order edge state is about 20 dB lower than that of the zeroth order edge state, it is still well within the detectable range under a reasonable signal-to-noise ratio.

Finally, we show that these edge states are remarkably robust to defects both in time and space. For instance, we consider a defect of size $\lambda_m/5$ in the vicinity of the interface located at *x*=0 between the two PCs, where the density is static. Thus, the whole system has its density following the form:

$$\rho = \rho_0\big(1 - (\alpha + \beta \sin(\omega_m t)).\cos(k_m x)\big) \quad \text{for } x < -\lambda_m/10\,,$$
$$\rho = \rho_0 \quad \text{for } -\lambda_m/10 < x < \lambda_m/10\,, \qquad (6)$$
$$\text{and} \quad \rho = \rho_0\big(1 + (\alpha + \beta \sin(\omega_m t)).\cos(k_m x)\big) \quad \text{for } x > \lambda_m/10\,.$$

where $\alpha = 0.5$ and $\beta = 0.2$.

We conduct a similar study to show that this defect does not affect the characteristic of the PC. Figure 4(b) shows the density variation during one modulation period $T_m = 2\pi/\omega_m$ where the defect can be observed as the constant density near *x*=0. Figure 4(a) presents the transmission spectrum for wave propagation through the structure where we can clearly see that the curve is very similar to the one of the system without the defect in Fig. 3(a). In fact, the two curves would virtually overlap if we were to plot them together. We can also observe the edge state inside this BG (the peak) at the normalized frequency $\Omega_0$=0.5 along with two transmission peaks located at 0.016 and 0.9841 which correspond to the excitation of the first order edge states outside the BG at $\Omega_0 \pm \Omega_m$. Because of the defect, these frequencies are slightly offset from the original ones (0.014, 0.04981 and 0.9821) by a small amount that is roughly 0.002. We plot in Fig. 4(c) the spatial-frequency spectrum of the wave field for monochromatic excitations at $\Omega_0 = 0.5$ and $\Omega_0 - \Omega_m = 0.016$, respectively in the left and right panels. We can clearly observe the same results as in Fig. 3(c) for the system without the defect. The edge states, including the anomalous ones, are therefore topologically protected against the defect and can be excited at a deep subwavelength scale where the incident wavelength is $\lambda \approx 62.5\, \lambda_m$ in this case.

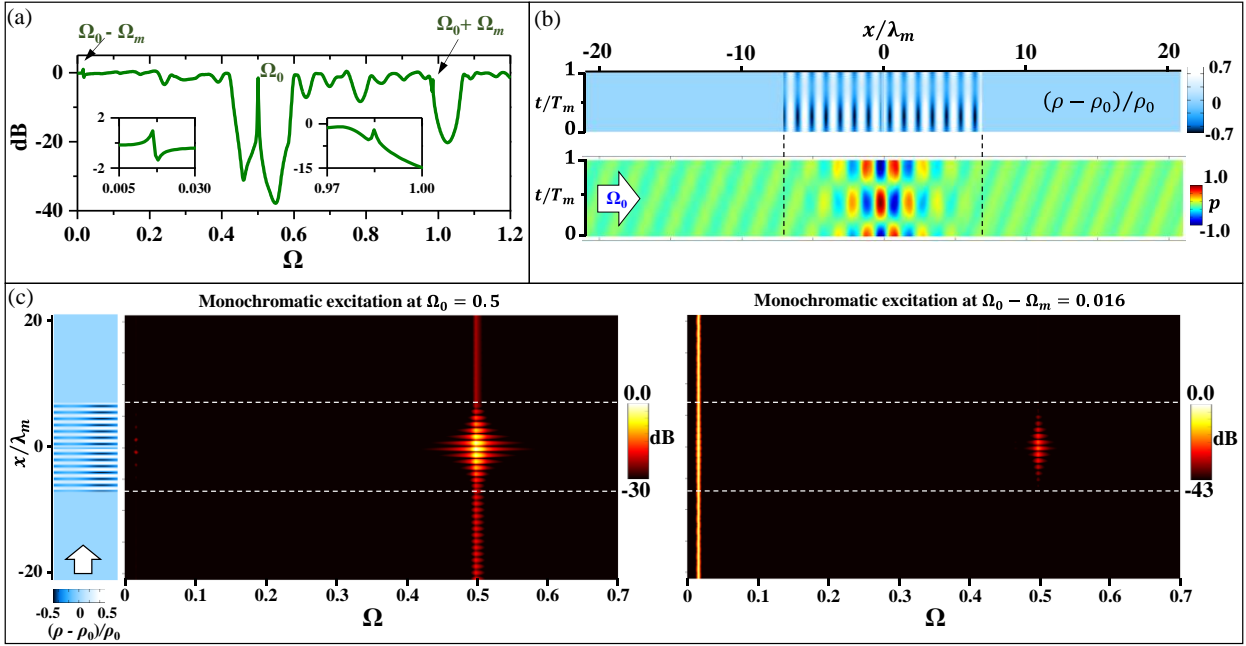

Fig.4. (a) Transmission spectrum through the space-decoupled time modulated system composed of two PCs and a static density defect in between, and a close-up view near $\Omega_0 \pm \Omega_m$ (inset figures). (b) (Top panel) Transmission configuration of the structure with density modulation of $\rho = \rho_0(1 - (\alpha + \beta sin(\omega_m t)).cos(k_m x))$ for $x < -\lambda_m/10$; $\rho = \rho_0$ for $-\lambda_m/10 < x < \lambda_m/10$; and $\rho = \rho_0(1 + (\alpha + \beta sin(\omega_m t)).cos(k_m x))$ for $x > \lambda_m/10$, presented within one modulation period $T_m$. (Bottom panel) Pressure field for incident wave at $\Omega_0 = 0.5$. (c) Spatial-frequency spectra of the wave field for incident wave excitations at $\Omega_0 = 0.5$ (left panel) and $\Omega_0 - \Omega_m = 0.016$ (right panel).

In conclusion, we propose to enrich the functionalities of topological devices via time modulation, which gives rise to the duplication of topological edge states across the entire spectrum. This feature can be used to achieve deep subwavelength manipulation of the edge state in the Bragg regime, which could be proven useful for designing miniature topological devices for wave guiding, sensing, and excitation, among other intriguing functionalities. Furthermore, the system presented here remains reciprocal while we invite the reader to the supplementary material [56] where we study the space-coupled time modulated system which breaks reciprocity that, in conjunction with the anomalous edge states, could engender exotic unidirectional wave behaviors. Future work will expand the model to higher dimensions where even richer physics can be foreseen. It would be also interesting to construct a real physical model to experimentally validate the theory. Active acoustic metamaterials [45] and piezoelectric materials [49,50] could be viable options in this regard.

**Acknowledgments**
The authors would like to thank Dr. Meng Xiao for his critical comments on the paper.